# Strong Correlation Driven Quadrupolar to Dipolar Exciton Transitions in a Trilayer Moiré Superlattice

Yuze Meng[1,#], Lei Ma[1,2,#], Li Yan[1,2,#], Ahmed Khalifa[1,#], Dongxue Chen[2,#], Shuai Zhang[1], Rounak Banerjee[3], Takashi Taniguchi[4], Kenji Watanabe[5], Seth Ariel Tongay[3], Benjamin Hunt[1], Shi-Zeng Lin[6,7], Wang Yao[8], Yong-Tao Cui[9,10], Shubhayu Chatterjee[1], Su-Fei Shi[1,*]

1. Department of Physics, Carnegie Mellon University; Pittsburgh, PA 15213, USA
2. Department of Chemical and Biological Engineering, Rensselaer Polytechnic Institute; Troy, NY 12180, USA
3. School for Engineering of Matter, Transport and Energy, Arizona State University; Tempe, AZ 85287, USA
4. Research Center for Materials Nanoarchitectonics, National Institute for Materials Science; 1-1 Namiki, Tsukuba 305-0044, Japan
5. Research Center for Electronic and Optical Materials, National Institute for Materials Science; 1-1 Namiki, Tsukuba 305-0044, Japan
6. Theoretical Division, T-4, Los Alamos National Laboratory; Los Alamos, New Mexico 87545, USA
7. Center for Integrated Nanotechnologies (CINT), Los Alamos National Laboratory; Los Alamos, New Mexico 87545, USA
8. Department of Physics, University of Hong Kong; Hong Kong, China
9. Department of Physics and Astronomy, University of California at Riverside; Riverside, CA 92521, USA
10. Department of Materials Science and Engineering, University of California at Riverside; Riverside, CA 92521, USA

[#] These authors contributed equally to this work
[*] Corresponding author: sufeis@andrew.cmu.edu

**Abstract**

**The additional layer degree of freedom in trilayer moiré superlattices of transition metal dichalcogenides enables the emergence of novel excitonic species, such as quadrupolar excitons, which exhibit unique excitonic interactions and hold promise for realizing intriguing excitonic phases and their quantum phase transitions. Concurrently, the presence of strong electronic correlations in moiré superlattices, as exemplified by the observations of Mott insulators and generalized Wigner crystals, offers a direct route to manipulate these new excitonic states and resulting collective excitonic phases. Here, we demonstrate that strong exciton-exciton and electron-exciton interactions, both stemming from robust electron correlations, can be harnessed to controllably drive transitions between quadrupolar and dipolar excitons. This is achieved by tuning either the exciton density or electrostatic doping in a trilayer semiconducting moiré superlattice. Our findings not only advance the fundamental understanding of quadrupolar excitons but also usher in new avenues for exploring and engineering many-body quantum phenomena through novel correlated excitons in semiconducting moiré systems.**

**Main Text**

Due to the enhanced light-matter interaction, monolayer transition metal dichalcogenides (TMDCs) provide an exciting platform for exploring excitons[1-3] and their many-particle complexes[4-6], which exhibit superior optical properties and a valley degree of freedom[7]. The van der Waals (vdW) assembly of TMDC heterobilayers enables long-lived interlayer excitons with large out-of-plane dipole moments that can be sensitively tuned via an external electric field[8,9]. Additionally, precise control of the twist angle or lattice mismatch gives rise to TMDC moiré superlattices that can further modify excitons in unprecedented ways[10-12]. Recently, it has been shown that excitons strongly interact with correlated electrons and also with each other, leading to exciting opportunities for realizing many-body quantum states of bosonic particles or even bosonic and fermionic particle mixtures[13-21].

The layer degree of freedom (DoF) is a powerful knob to help explore and engineer electronic and excitonic states via the layer-spin-valley locking[3,22-25]. The added layer DoF can be utilized to construct new types of excitons, such as quadrupolar excitons (QXs) with distinct excitonic interactions that can be employed to realize novel quantum phase transitions[26-34]. However, the ability to control and manipulate QX states remains to be demonstrated. Strong electronic correlations in TMDC moiré superlattices present new opportunities to achieve the desired control. Here, we show that such electronic correlations can be leveraged to control excitonic states by tuning the exciton-exciton or electron-exciton interaction. Further, we illustrate the crucial role played by the moiré potential, often overlooked while modeling QXs, in enhancing their interactions.

In this work, we construct a dual-gated trilayer moiré superlattice device made of angle-aligned trilayer: $WSe_2/WS_2/WSe_2$, which has been shown previously[27] to host QXs with distinctly different properties from dipolar interlayer excitons. We demonstrate that the strong exciton-exciton and exciton-electron interaction, both originating from the strong electron correlation, critically affect the properties of QXs and can be utilized to drive QX-to-dipolar exciton (DX) transitions, thereby substantially changing the nature of the excitonic interactions. By increasing the optical excitation intensity that increases the exciton density, we first realize a QX-to-DX transition at intermediate excitation intensity and subsequently demonstrate a second transition from DX-to-QX at higher excitation intensity. The former occurs when two excitons occupy the same moiré site, and the exciton-exciton repulsion renders the two oppositely oriented DXs the energetically favorable configuration. The latter occurs when the exciton repulsion at the increased exciton density frees excitons from the moiré potential confinement, and the ground state is a QX liquid in which the kinetic energy is not negligible. We can also control the QX-to-DX transition by controlling the doping of the moiré trilayer of $WSe_2/WS_2/WSe_2$. If the trilayer moiré superlattice is at the hybridized Mott insulator state, i.e., one hole per moiré unit cell shared by the top $WSe_2$ (t-$WSe_2$) and bottom $WSe_2$ (b-$WSe_2$) but laterally confined in one moiré cell, adding more holes through electrostatic gating will drive the QX-to-DX transition. In contrast, any noticeable addition of electrons electrostatically will induce the QX-to-DX transition. Our work demonstrates the ability to control excitonic phases via two distinct tuning knobs by leveraging strong correlations in TMDC moiré superlattices.

A typical dual-gated device is schematically shown in Fig. 1a, with the electric field and doping independently controlled via the combination of top and back gate voltages. The schematic for the trilayer moiré structure, $WSe_2/WS_2/WSe_2$, is shown in Fig.1b. As reported previously[26,27], the $WSe_2/WS_2/WSe_2$ moiré trilayer host QXs, a result of the hybridization of the valence bands of the top (t-$WSe_2$) and bottom $WSe_2$ (b-$WSe_2$) layers (Fig. 1c). The QX is the energetically favorable configuration as the energy is lowered by $\Delta_{DQ}$, compared with that of DX (Fig. 1c). The existence of QXs in this moiré trilayer is evident in the electric-field-dependent PL spectra of the moiré trilayer device near the intrinsic regime (Fig. 1d), which exhibits characteristic quadratic electric field dependence due to the Stark effect, distinctively different from the linear Stark shifts for the dipolar interlayer exciton in the $WS_2/WSe_2$ moiré bilayer regions. It is interesting that the PL intensity from the moiré trilayer region shows three intensity resonances at different electric fields: one at zero electric field and two symmetric ones at a nominal electric field about ±30 mV/nm (see Method about nominal electric field calculations). It is worth noting that when the electric field magnitude exceeds ~30 mV/nm, the shift of PL peak positions can be well described by the linear Stark shift (dashed red lines in Fig. 1d). Therefore, we assign the central PL resonance at the zero electric field as QX and the two resonances at around ±30 mV/nm as dominated by DXs.

We have performed optical excitation intensity dependence measurements to investigate the three distinct PL resonances. With the optical excitation of a continuous wave (CW) laser centered at photon energy 1.70 eV, we find that the three PL resonances evolve differently as we increase the excitation intensity, which increases the exciton density. As we increase the excitation intensity from 2.6 µW/µm² (Fig. 2a) to 6.4 µW/µm² (Fig. 2b), the resonance at the zero electric field from the QX PL is much less pronounced compared to that of DXs at the electric field of about ±30 mV/nm. However, at a further increased excitation intensity of 127.4 µW/µm² (Fig. 2c), the QX PL becomes the dominant one. A detailed excitation intensity dependence study can be found in Extended Data Fig. 2.

Assuming that the quantum yield ratio between the QXs and DXs is not a sensitive function of the excitation intensity (details in Supplementary Section 5), the relative PL intensity of the resonance at zero electric field and ~ ±30 mV/nm corresponds to the density ratio of the QXs to DXs. For better illustration, we normalize the PL spectra for each excitation intensity by dividing the integrated PL intensity at different electric fields by the respective maximum value, shown as the color plot in Fig. 2d. The maximum value of 1 at each excitation intensity reveals the dominating exciton species under this excitation condition.

The normalized PL intensity as a function of both applied electric field and excitation intensity exhibits three distinct regions, as shown in Fig. 2d. Region I corresponds to a small excitation intensity below 3.2 µW/µm², (white dashed line in Fig. 2d), in which the QX PL at zero electric field dominates (see also Extended Data Fig.1a, d). In Region II, the QX resonance is less pronounced than that of DXs. There is some asymmetry between the two DX resonance intensities, likely due to the inevitable asymmetric coupling during fabrication. In Region III, at an elevated excitation intensity beyond 88.9 µW/µm² (white dashed line in Fig. 2d), the QX resonance becomes the dominant one again. Fig. 2a-c shows the characteristic electric field dependence of the PL spectra for the trilayer moiré device in the corresponding Regions I, II, and III, respectively. The three distinct regions have been reproduced in 5 devices we have studied (such as the one shown in Extended Data Fig. 1).

A consistent explanation of all the above experimental observations can be obtained by a simple model incorporating two basic ingredients: (i) the confining moiré potential and (ii) strong exciton-exciton interactions[30]. Under a small excitation intensity, we have a density of less than one exciton per moiré cell, and the exciton is trapped in one of the high symmetry points within the moiré unit cell with a large moiré potential confinement. The hole is shared between the t- and b-$WSe_2$ layers due to the hybridization of valence bands, forming the QX state corresponding to the PL intensity resonance at zero electric field. Under a sufficiently large electric field, the valence band maximum (VBM) degeneracy of the t- and b-$WSe_2$ layers is lifted, and the hole is polarized to either the t- or b-$WSe_2$ layer, resulting in the DX state. The breakdown electric field magnitude of about 30 mV/nm, together with the t- and b-$WSe_2$ layer separation of ~1.4 nm, suggests an energy shift of about 42 meV between the t- and b-$WSe_2$ layers, which is about twice the hybridization energy of 22 meV extracted for this device through the two-level hybridization fitting (details in Supplementary Section 3).

At the increased excitation intensity that corresponds to a density of two excitons per moiré cell, the scenario is drastically different. Due to the deep moiré potential confinement of more than 100 meV[18,21], the two excitons are expected to be confined in the same potential well, consistent with our previous report[18]. As a result, the effective exciton spacing is much reduced compared to the scenario in the absence of any moiré potential confinement. Consequently, the Coulomb interaction leads to the ground state of two oppositely oriented DXs over the two QXs[30], schematically shown in Fig. 2e(the middle panel). To demonstrate this effect quantitatively, we model the interacting excitons at each moiré site by considering the harmonic confinement potential obtained via an expansion of the full moiré superlattice potential, and the Coulomb interaction of the holes in the top/bottom layer and the electrons in the middle layer. Specifically, we consider the energetics of two distinct configurations: (i) two oppositely oriented DXs and (ii) two quadrupolar excitons. Following variational optimization of the energy with the inter-exciton separation in each case, the configuration with two DXs is shown to have lower energy, supporting our interpretation (see Supplementary Section 10 for details). Under this condition, the QX PL at zero electric field is relatively reduced (Fig. 2b and Region II of Fig. 2d). We estimate that, for the excitation photon energy of 1.70 eV, a density of two excitons per moiré cell corresponds to the excitation intensity of about 5.2 µW/µm² (see

Supplementary Section 4 for calculation details), which is in good agreement of the Region I to Region II transition threshold of about 3.2 µW/µm².

At a further increased excitation intensity, i.e., much higher exciton density, the repulsive exciton interaction overcomes the moiré potential, and most excitons will no longer be confined at the moiré site. This interpretation is supported by the blueshift of about 20 meV for the PL at zero field from Fig. 2a to Fig. 2c. As a result, the PL is dominated by the unconfined excitons with reduced Coulomb interactions (increased exciton spacing), which again reside in the energetically favorable QX state due to the larger energy gain of hybridization $\Delta_{DQ}$.

We also investigate the effect of electron-exciton interaction on the formation of QXs by performing electric-field-dependent PL spectroscopy studies for different electrostatic doping levels, as presented in Fig. 3. Here, we normalize the doping density with respect to the moiré superlattice as the filling factor n, with n=-1.5 corresponding to 1.5 holes per moiré unit cell and n~0.3 corresponding to about 0.3 electrons per moiré unit cell. As we increase the hole doping from n=0 (Fig. 3c) to n=-0.7 (Fig. 3b) and further to n=-1.5 (Fig.3a), it is evident that the three PL intensity resonances evolve as a function of the p-doping: at n=-0.7 only the QX resonance remains, while at n=-1.5 only the two symmetric DX resonances are pronounced at electric fields around ±30 mV/nm.

It is interesting to note that as we dope the device to the electron side (Fig. 3d, e), the PL peak position becomes a linear function of the electric field (Fig. 3f, green and black dots), indicating that the PL stems from DXs instead of QXs. A detailed PL evolution as a function of electron doping can be found in Supplementary Section 8. It suggests that any noticeable addition of electrons to the system would convert QXs to DXs, which is also supported by the peak position phase diagram in Supplementary Section 9. In the p-doping region, the PL peak position is less sensitive to the nominal electric field, as the electrostatically introduced holes on the t- and b-WSe$_2$ would screen the effective field on the excitons (detailed discussion in Supplementary Section 13).

To better understand the QX to DX transitions, we investigate the integrated interlayer exciton PL intensity as a function of the filling factor (doping) and the nominal electric field in Fig. 4a. The data exhibits a rich phase diagram in the hole-doping regime that can be separated into multiple regions by the dashed lines.

First, the most pronounced feature is the strong PL between the filling factor 0 and -1 at zero electric field, annotated by the black dashed line. The PL in this range is of the QX nature, evidenced by its quadratic electric field dependence (Fig. 3b). As the p-doping increases further to beyond -1, the QX PL is quenched, suggesting a QX-to-DX transition due to strong electron correlation. At n=-1, one hole is localized laterally on a moiré site, forming a hybridized Mott insulator[27]. The localized hole at each moiré site coexists with the optically excited exciton, which induces a repulsive interaction in a way similar to the exciton-exciton interaction in the case of double-exciton occupancy we discussed previously (Fig. 2e middle panel). Instead of forming a QX that requires efficient hole tunneling between t- and b-WSe$_2$ layers, the repulsion renders the DX an energy-favorable configuration, with the optically excited hole separated from the electrostatically induced hole that is localized in the other WSe$_2$ layer (see Supplementary Section 10 for details).

Second, near zero doping, the PL intensity is strong along the vertical white dashed line in Fig. 4a. At electric fields beyond ±30 mV/nm, the QXs transition to DXs, which is the scenario we have discussed in Fig. 1d. The DX is polarized to either the t-WSe$_2$/WS$_2$ bilayer or the b-WSe$_2$/WS$_2$ bilayer.

Third, the electric-field-driven QX-to-DX transition is also evident at around ±30 mV/nm at n= -1 (points A and B in Fig. 4a). Upon further hole doping, the DX PL intensity quickly decreases as the screening by the surrounding electrostatically introduced holes increases. The DX PL evolution as a function of doping and electric field thus reveals the free carrier distribution around DX. The orange dashed lines originating from point A or B beyond n=-1 track the local maximum of DX PL intensity. These lines run approximately parallel with either the top or bottom gate axis, corresponding to varying the hole doping in either t- or b-WSe$_2$ layer. This behavior is in excellent agreement with a model that considers the quantum capacitance of each TMDC layer (details in Supplementary Section 9).

Finally, the PL is much weaker in the electron-doping region where the QXs transition to DX (Fig.4a). Extended Data Fig. 4 plots the PL of the n-doping region under increased optical excitation intensity. The PL peak position change as a function of the doping and electric field is consistent with the expectation for DXs (Supplementary Section 9), further confirming the QX-to-DX transition in the electron-doping region. The sensitive dependence of the nature of the exciton on electron doping arises from the fact that the electrostatically introduced electrons reside in the middle WS$_2$ layer. To introduce electron doping, the electric fields from both gates attract electrons to occupy the WS$_2$ layer in the middle by lowering the chemical potential for electrons. Accordingly, the chemical potential for holes will increase, which will suppress the hole probability in the middle layer. Considering that the original QX ground state is the bonding state with a finite spectral weight of the QX hole wavefunction in the middle layer[30], the energy of the QX in the n-doping regime will increase, potentially yielding the DX as the lower energy state (see Supplementary Section 11 for details).

In summary, we have successfully demonstrated controllable transitions between QXs and DXs by tuning the exciton density or electrostatic doping in the WSe$_2$/WS$_2$/WSe$_2$ moiré trilayer device. The unprecedented control of excitonic interactions realized via these transitions is based on our improved understanding of the effects of the strong electron-exciton correlation and the moiré potential confinement on QXs. Our study paves the way for engineering new types of correlated excitons in multilayer moiré systems, ushering in new avenues for investigating emergent many-body quantum phenomena.

**Acknowledgments**

We thank Chenhao Jin for the helpful discussions. S.-F.S. acknowledges the support from NSF (DMR-1945420, DMR-2428545, ECCS-2344658, and ECCS-2139692) and a community collaboration award from the Pittsburgh Quantum Institute. The optical spectroscopy measurements were supported by an AFOSR DURIP award through Grant FA9550-23-1-0084. Y.-T.C. acknowledges support from NSF under awards DMR-2104805 and DMR-2145735. K.W. and T.T. acknowledge support from the JSPS KAKENHI (Grant Numbers 21H05233 and 23H02052) and World Premier International Research Center Initiative (WPI), MEXT, Japan. S.T. acknowledges primary support from DOE-SC0020653 (materials synthesis), Applied Materials Inc., DMR 2111812, DMR 2206987, and CMMI 2129412. The work at LANL is partially supported by the U.S. Department of Energy (DOE) National Nuclear Security Administration (NNSA) under Contract No. 89233218CNA000001 through the Laboratory Directed Research and Development (LDRD) Program and was performed, in part, at the Center for Integrated Nanotechnologies, an Office of Science User Facility operated for the DOE Office of Science, under user proposals No. 2018BU0010 and No. 2018BU0083.

## Author contributions

S.-F.S. conceived the project. Y.M. and D.C. fabricated heterostructure devices. L.Y. and L.M. performed the optical spectroscopy measurements. R.B. and S.T. grew the TMDC crystals. T.T. and K.W. grew the BN crystals. S.L. and W.Y. contributed to the theoretical understanding. A. K. and S.C. contributed to the theoretical calculations. S.-F.S, Y. M., L.M., and L.Y. analyzed the data. S.-F.S. wrote the manuscript with the help of Y.-T.C., S.L., L.Y., L.M., Y.M., S.Z., B.H., and input from all authors.

These authors contributed equally: Yuze Meng, Lei Ma, Li Yan, Ahmed Khalifa, and Dongxue Chen.

## Competing interests

The authors declare no competing interest.

# Figures

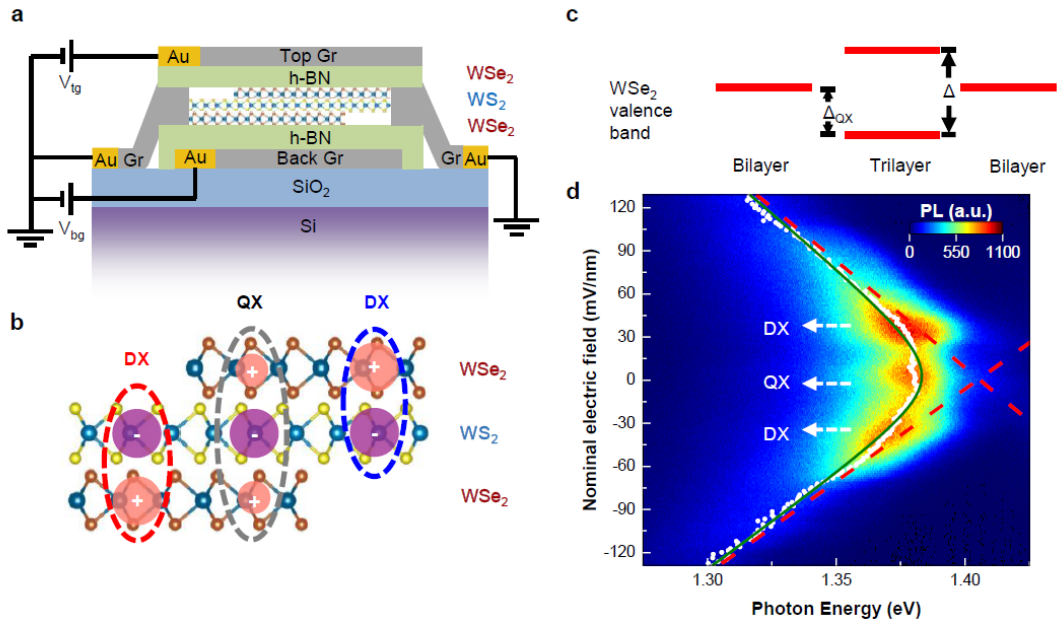

**Fig.1 | QXs in the WSe$_2$/WS$_2$/WSe$_2$ moiré trilayer.** (a) Schematic of the WSe$_2$/WS$_2$/WSe$_2$ moiré trilayer device. (b) Schematic of the DXs and QXs in the moiré bilayer and trilayer regions. (c) Schematic of the band hybridization. (d) PL spectra of QXs from the trilayer region as a function of the nominal electric field. All the data in this manuscript is taken with a CW laser centered at 1.70 eV as the excitation source. The sample temperature is kept at 7 K. The excitation intensity is 0.8 µW/µm$^2$ for Fig. 1. White dots are extracted PL peak positions. The green line is the quadratic fitting of the peak positions. The red dashed line is the linear fitting of the peak position at the high electric field.

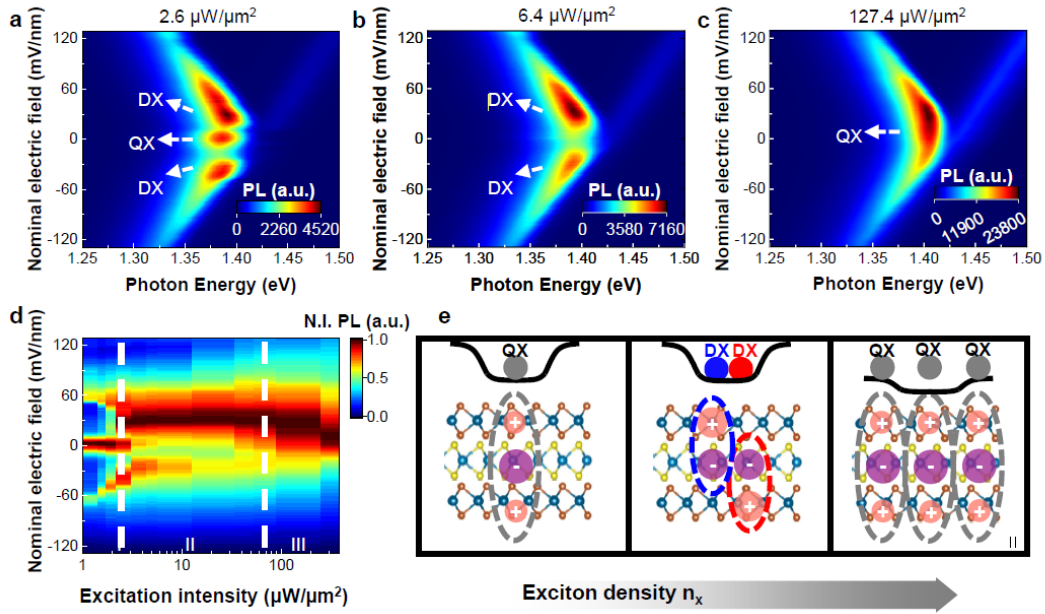

**Fig.2 | Exciton-density-driven transitions between QXs and DXs.** (a)-(c) PL spectra as a function of the nominal electric field for the CW excitation centered at 1.70 eV, with the excitation intensity of 2.6 µW/µm², 6.4 µW/µm², and 127.4 µW/µm², respectively. The sample temperature is kept at 3.6 K. (d) The color plot of the normalized integrated PL intensity. For each excitation intensity in (d), the integrated PL intensity as a function of the electric field is normalized by the maximum integrated PL intensity. The white dashed lines label the excitation intensity of 3.2 µW/µm² and 88.9 µW/µm², which mark the threshold for Regions I, II, and III. (e) Schematics of QXs and DXs configurations for different exciton densities in Regions I, II, and III, respectively.

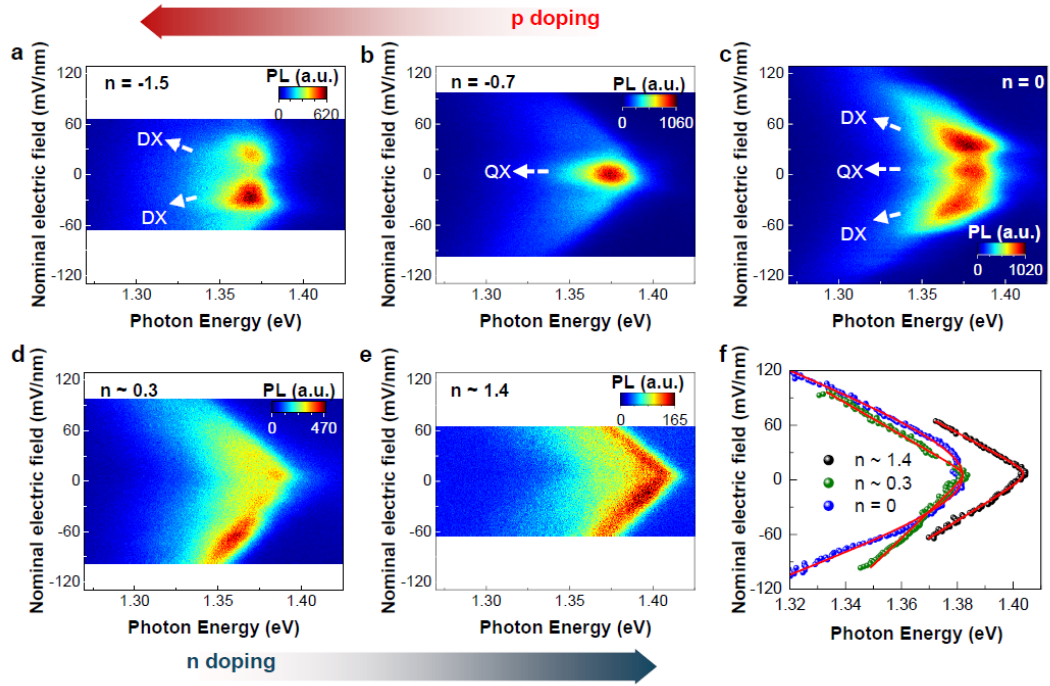

**Fig.3 | Electrostatic-doping-driven QXs to DXs transitions.** (a)-(e) are electric field-dependent PL spectra with filling factor of n=-1.5, -0.7, 0, ~0.3 and 1.4, respectively. The excitation intensity is 0.8 µW/µm$^2$, which corresponds to an estimated exciton density of 0.3 exciton per moiré cell. The sample temperature is 7 K. (a) and (b) are in the hole doping regime. (c) is in the intrinsic regime, while (d) and (e) are in the electron doping regime. (f) PL peak positions extracted for n=0 (blue dots), 0.3 (green dot), and 1.4 (blue dot), respectively. The red lines are the fittings by the QX model at the filling factor n=0 (c), n~0.3 (d), and n~1.4 (e), respectively.

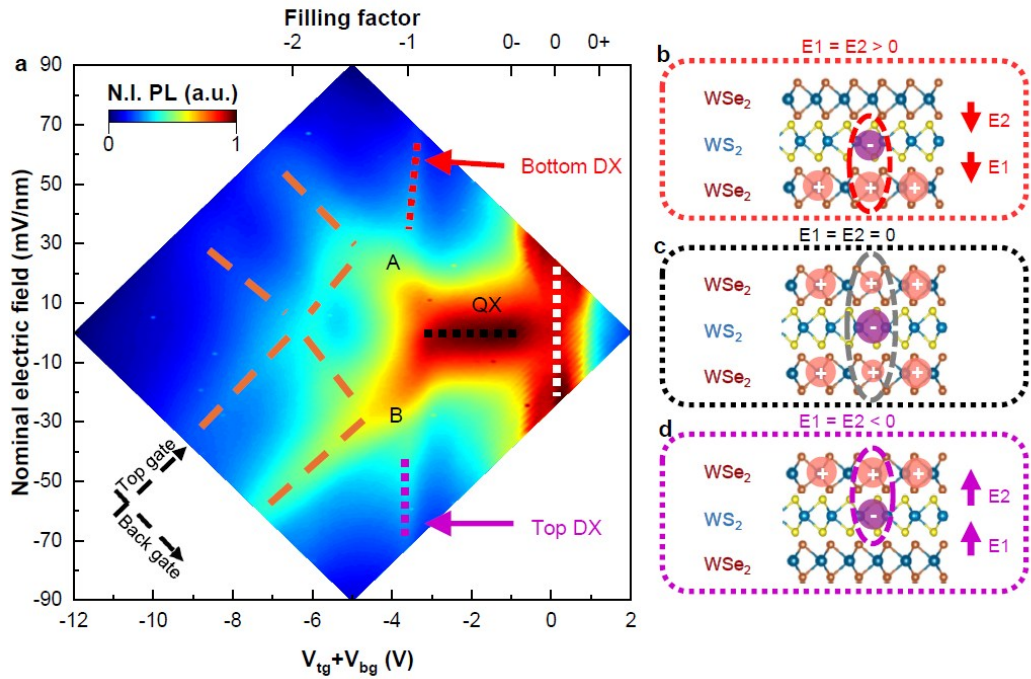

**Fig.4 | Phase diagram revealing QX to DX transition.** (a) Integrated PL intensity at 7 K as a function of doping and nominal electric field with an excitation intensity of 1.3 µW/µm$^2$. Estimated exciton density is 0.5 exciton per moiré cell (b)-(d) show the doping distribution and interlayer electric field direction of the trilayer region along the red, black, and purple dot-dashed line, respectively.

**Extended Figures:**

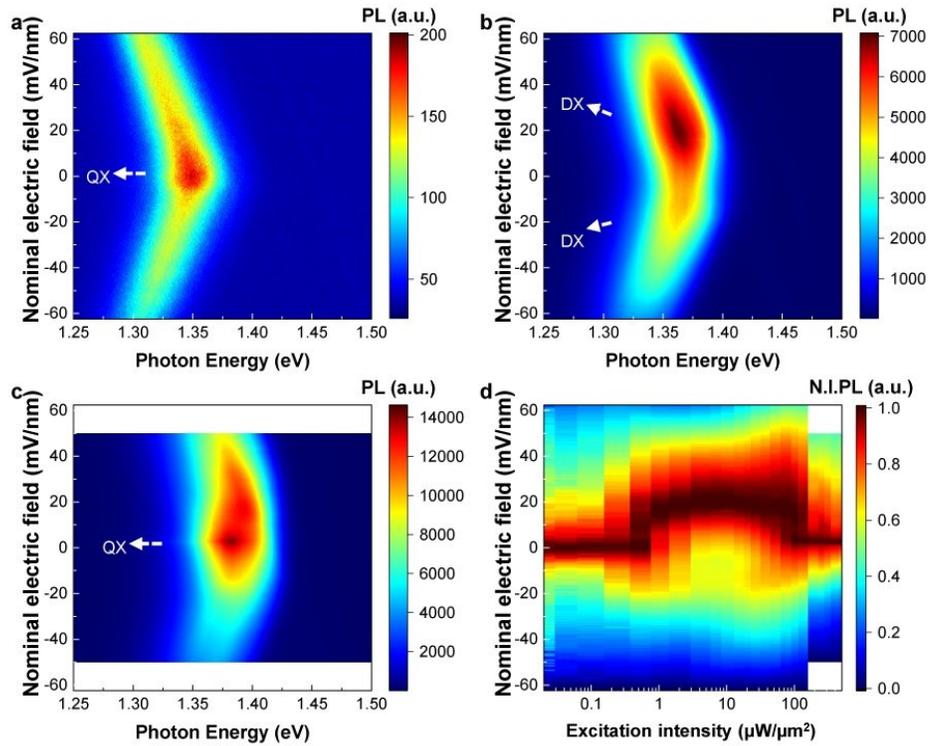

**Extended Data Fig.1 | Electric field dependent PL spectra for device D2.** (a), (b), and (c) are PL spectra acquired under the CW optical excitation with the excitation intensity of 0.053 µW/µm$^2$, 12.7 µW/µm$^2$, and 259.1 µW/µm$^2$, respectively. (d) Colour plot of the normalized integrated PL intensity as a function of the excitation intensity and nominal electric field. The measurements were performed at a temperature of 3.6 K.

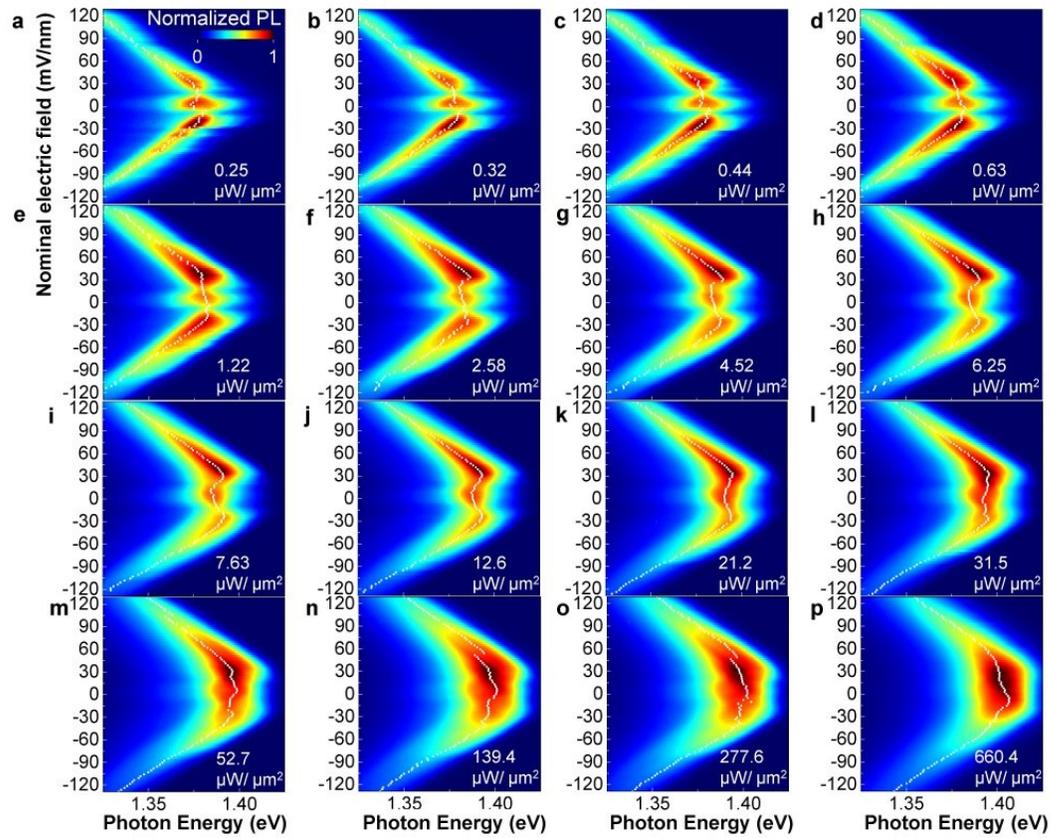

**Extended Data Fig.2 | Electric field dependent PL spectra at different excitation intensities.** The extracted PL peak positions are shown as white dots. The measurements were performed at a temperature of 3.6 K.

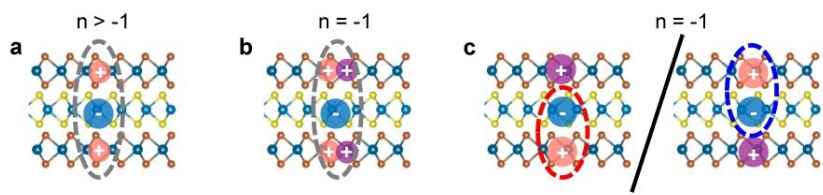

**Extended Data Fig.3 | Schematics of QX-to-DX transition at n=-1.** The pink and purple spheres refer to the optically excited hole and electrostatically introduced hole, respectively. (a) shows the exciton configuration as a QX for the doping of less than one hole per moiré unit cell. (b) and (c) show two competing exciton configurations when electrostatic doping at n=-1. Calculations in Supplementary Section 10 show that the configuration of (c) is of the lower energy.

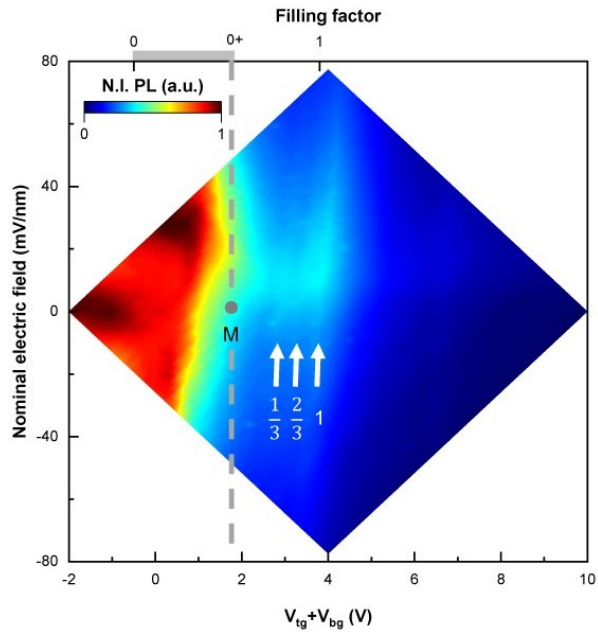

**Extended Data Fig.4 | Integrated PL intensity as a function of the doping and nominal electric field in the n-doping regime.** The gray bar indicates the intrinsic region of the device, and a detailed discussion can be found in Supplementary Section 9. White arrows suggest the filling factor of n=1/3, 2/3, and 1, respectively. The PL spectra were taken with the CW excitation intensity of 3.2 µW/µm$^2$ and temperature of 7 K.

**Method**

**Sample fabrication**

The dry pick-up method (described in our earlier work[35]) is applied to fabricate the TMDC moiré trilayer heterostructures. The whole stack is then melted on the substrates with pre-pattern electrodes at 170 °C and rinsed with chloroform and isopropanol sequentially to remove polycarbonate (PC) residues. The device was then dried with nitrogen gas and annealed in a vacuum (< $10^{-6}$ torr) at 250 °C for 8 hours.

**Optical characterizations**

For all measurements except for the PL measurements in Fig. 2, the moiré trilayer devices were mounted in a close-cycle optical cryostat (Montana Instruments). PL measurements in Fig. 2 were taken with the same device but in an Attodry 1000 system. A home-built confocal microscope system was used to focus the laser on the sample. The PL signals were collected through a spectrometer (Princeton Instruments) equipped with a silicon charge-coupled device (Princeton Instruments). The PL measurements in the main text were performed with a CW diode laser centered at 1.70 eV. The reflectance contrast measurements were performed with a super-continuum laser (YSL Photonics) with a 10 MHz repetition rate.

The polarized second harmonic generation (SHG) measurements were performed with a pulsed laser excitation centered at 1.38 eV (Ti: Sapphire, Coherent Chameleon) with a repetition rate of 80 MHz and a power of 40 mW. During the measurements, the sample was mounted in a vacuum dewar with pressure < 10⁻⁶ torr. The crystal axes of the sample were fixed. The polarized SHG signal was analyzed using a half-waveplate and a polarizer.

**Determination of the nominal electric field**

We obtained the thickness of the top and bottom h-BN flakes by atomic force microscopy (AFM) and defined them as $d_1$ and $d_2$. The nominal electric field is defined as the electric field in trilayer heterostructures, which is given by $E = \frac{\epsilon_{BN}}{2\epsilon_{TMDC}}\left(\frac{V_{TG}}{d_1} - \frac{V_{BG}}{d_2}\right)$, where $V_{TG}$ ($V_{BG}$) is the top (bottom) gate voltage, $d_1$ ($d_2$) is the thickness of the top (bottom) BN flake determined by atomic force microscopy, $\epsilon_{BN} = 3.9$ and $\epsilon_{TMDC} = 7.2$ are the relative dielectric constants of h-BN and TMDC, respectively[9].

**Data Availability**

All data that support the plots within this paper and other findings of this study are available from the corresponding authors upon reasonable request.

**Method-Only References**